\documentclass[journal]{IEEEtran}

\usepackage[utf8]{inputenc}
\usepackage[T1]{fontenc}
\usepackage{siunitx}
\usepackage[colorlinks=true, allcolors=black]{hyperref}
\usepackage{amsmath}
\usepackage{amsfonts}
\usepackage{soul} 
\usepackage{booktabs} 
\usepackage{adjustbox}
\usepackage{subcaption}
\usepackage{todonotes}
\usepackage{makecell}
\usepackage{pgfplots}
\usetikzlibrary{patterns}
\usepackage{multirow}
\usepackage{color, colortbl}
\usepackage{pgf-pie}
\usepackage{tikz,xcolor}
\definecolor{lime}{HTML}{A6CE39}
\DeclareRobustCommand{\orcidicon}
{
    \begin{tikzpicture}
    \draw[lime, fill=lime] (0,0) circle [radius=0.16] 
    node[white] {{\fontfamily{qag}\selectfont \tiny ID}};    \draw[white, fill=white] (-0.0625,0.095) circle [radius=0.007];    
    \end{tikzpicture}
    \hspace{0mm}}
\foreach \x in {A, ..., Z}{%
    \expandafter\xdef\csname orcid\x\endcsname{\noexpand\href{https://orcid.org/\csname orcidauthor\x\endcsname}{\noexpand\orcidicon}}
}

\begin{document}

\title{A Virtual Reality Training System for Automotive Engines Assembly and Disassembly}

\author{Gongjin Lan$^*$\hspace{-1mm}\orcidA{}\hspace{-1mm}, \IEEEmembership{Member, IEEE}, Qiangqiang Lai, Bing Bai, Zirui Zhao, Qi Hao$^*$\hspace{-1mm}\orcidB{}\hspace{-1mm}, \IEEEmembership{Member, IEEE}
\thanks{This work was partially supported by the Shenzhen Fundamental Research Program (JCYJ20200109141622964, JCYJ20220818103006012), the National Natural Science Foundation of China (62261160654), the GuangDong Basic and Applied Basic Research Foundation (2021A1515110641). (Corresponding author: Gongjin Lan; Qi Hao.)}
\thanks{Gongjin Lan, Qiangqiang Lai are with the Department of Computer Science and Engineering, Southern University of Science and Technology, Shenzhen, 518055, China (e-mail: langj@sustech.edu.cn, laiqq@mail.sustech.edu.cn)}
\thanks{Bing Bai, Ziru Zhao, Qi Hao are with the Department of Computer Science and Engineering, Research Institute of Trustworthy Autonomous Systems, Southern University of Science and Technology, Shenzhen, 518055, China (e-mail: \{11510163, zhaozr\}@mail.sustech.edu.cn, hao.q@sustech.edu.cn)}
}

\markboth{IEEE TRANSACTIONS ON LEARNING TECHNOLOGIES, VOL. XX, NO. XX, XXXX
}%
{Shell \MakeLowercase{\textit{et al.}}: Bare Demo of IEEEtran.cls for IEEE Journals}

\maketitle

\begin{abstract}
Automotive engine assembly and disassembly are common and crucial programs in the automotive industry.
Traditional education trains students to learn automotive engine assembly and disassembly in lecture courses and then to operate with physical engines, which are generally low effectiveness and high cost. 
In this work, we developed a multi-layer structured Virtual Reality (VR) system 
to provide students with training in automotive engine (Buick Verano) assembly and disassembly.
We designed the VR training system with 
The VR training system is designed to have several major features, including replaceable engine parts and reusable tools, friendly user interfaces and guidance, and bottom-up designed multi-layer architecture, which can be extended to various engine models.
The VR system is evaluated with controlled experiments of two groups of students.
The results demonstrate that our VR training system provides remarkable usability in terms of effectiveness and efficiency.
Currently, our VR system has been demonstrated and employed in the courses of Chinese colleges to train students in automotive engine assembly and disassembly.
A free-to-use executable file (Microsoft Windows) and open-source code are available at \url{https://github.com/LadissonLai/SUSTech_VREngine} for facilitating the development of VR systems in the automotive industry.
Finally, a video describing the operations in our VR training system is available at \url{https://youtu.be/yZe4YTwwAC4}.
\end{abstract}


%
\IEEEpeerreviewmaketitle

\vspace{-3pt}
\section{Introduction}
Virtual Reality (VR) is a type of human-machine interaction technology that simulates real-world environments, which has been applied to various domains, including entertainment (e.g., video games \cite{lan2016development}), education (e.g., medical \cite{bowyer2008immersive} or industrial training) and business (e.g., virtual meetings \cite{lan2016development2}). 
VR systems offer more effective learning with a lower cost and efficient way of learning, compared to traditional learning in lecture courses, and allow more training repetitions, especially when dealing with costly, rare, or dangerous environments \cite{langley2016establishing}.
These benefits apply to both the general population as well as specialists who need training for unique tasks, such as the automotive engine assembly and disassembly in this work.


In traditional automotive industry education, students usually receive theoretical training in lecture courses followed by practical operation with the real automotive engine under supervision, in which students generally lack sufficient motivation to further knowledge and skills 
\cite{martin2015augmented}.
Instead, students often feel much stress at the practical operation stage which could lead to mistakes that are harmful not only to equipment and even themselves.
In contrast, VR training immerses students in a three-dimensional space, simulating a real working environment, allowing them to practice skills in a safe environment.
In the VR system, students feel confident and can repeat the learning as much as necessary until they have successfully acquired the skills.
In addition, virtual reality enables fewer supervisors to teach more students in a safe environment.


Currently, many automotive companies, including Volkswagen and BMW, use VR to train workers on the automotive assembly line. 
With VR systems, automakers create a controlled and safe learning environment for reducing occupational injuries and increasing the effectiveness of training.
For example, 
Volkswagen saves costs by using VR systems for assembly training, as trainees do not need to visit the internal showrooms with real vehicles \footnote{\url{https://www.vrowl.io/volkswagen-assembly-vr-training/}}. 
New employees do not need an expensive real pre-production model at the beginning of the training and can even practice vehicle assembly at their homes or offices. 
In addition, Ford Motor Company has reduced its production line injury rate by 70\% since 2003 by using immersive VR training \footnote{\url{https://media.ford.com/content/fordmedia/fna/us/en/news/2015/07/16/ford-reduces-production-line-injury-rate-by-70-percent.html}}.
However, current VR systems in the automotive industry are proprietary and applied to the vehicle assembly line rather than the automotive engine assembly and disassembly.


The goal of this work is to provide students with a VR system for safe, efficient, and low-cost training in automotive engine assembly and disassembly. 
In order to develop such a VR system, we need to address several challenges: 1) a large number of models of engine parts and tools, 2) the development for smooth performance with friendly user interfaces and high frame rates to prevent dizziness, 3) the system optimization for low computing cost on common computer hardware.
In this work, we develop a multi-layer structured VR system that is a collaborative project with the partner of FXB Ltd. \footnote{\url{http://www.fengxb.com/}} to provide students with training in automotive engine (Buick Verano) assembly and disassembly.
Specifically, we modularly design the VR system by using Unity3D and VRTK plugins.
The VR system contains replaceable engine parts and reusable tools, friendly user interfaces and guidance, and bottom-up designed multi-layer architecture, which can be extended to various engine models.
Furthermore, we optimize the VR system to reach a frame rate of 90 fps for a smooth and non-vertiginous user experience. 
Our VR system offers many advantages in automotive engine assembly and disassembly: 1) improves the quality of student training and decreases the number of trainee errors; 2) provides efficient learning with less time for supervisors and students; 3) provides a lower cost training than the real training on physical engine; 4) enables the repeatable VR training as often as necessary; 5) provides a safe learning environment for reducing occupational injuries and machine damages.

We implemented and evaluated the VR system 
in a course taught in many vocational colleges in China
for training students.
The students were divided into two groups for controlled experiments: the VR group and the traditional group where students learn by listening to lectures.
The results show that our VR training system significantly improves student learning 
compared to the traditional training in the lecture and the training on physical engines.

The main contributions of our VR training system in automotive engine assembly and disassembly are summarized as:
\begin{enumerate}
    \item A VR system that provides efficient, safe, and low-cost training for quality students training and reduces trainee errors, occupational injuries and machine damage.
    \item The implementation and demonstration of the VR system in the courses of Chinese colleges.
    \item The free-to-use executable file (Microsoft Windows) and open-source code to promote VR development in the automotive industry.
\end{enumerate}

The rest of this paper is organized as follows. 
In \autoref{sec:related_work}, we review the existing studies that applied VR to the automotive industry. 
We present the design of our VR system and hardware setup in \autoref{sec:methodology}.
\autoref{sec:system_optimization} presents the system testing and optimization of our VR system.
The evaluation and results are presented in \autoref{sec:evaluation_results}.
Finally, we provide an in-depth discussion and conclusion in \autoref{sec:discussion_conclusion}.

\section{Related Work}
\label{sec:related_work}

VR technologies have been applied to various fields, such as first responder training, medical training, military training, and education \cite{xie2021review}.
Many studies apply VR to train first responders \cite{conges2020crisis,koutitas2021performance} (including fire fighters \cite{heirman2020exploring}) to reduce training costs and improve performance.
VR has been applied to medical training, such as the usage of VR games for Amblyopia treatment \cite{hurd2019virtual}, the incorporation of tactile sensation into the therapy session \cite{willis2019visual}, healthcare professionals to assess patients who show neurological symptoms of a stroke \cite{daher2018physical}.
Several studies investigate the applications of VR in military training, such as the simulation of different conditions of environments \cite{rushmeier2019content}, 
and the training of army artillery \cite{girardi2019virtual}, pilots training in a live aircraft \cite{dalladaku2020assessing}.
Furthermore, VR offers a number of benefits for education.
Kaminska et al. \cite{kaminska2019virtual} reviewed the studies of VR and its applications in education areas such as engineering \cite{soliman2021application} and health-related education \cite{pottle2019virtual,chen2020effectiveness}.
Although these studies apply VR to common applications, there is a lack of studies that apply VR to the assembly and disassembly training for automotive engines.

There are a few studies that apply VR to assembly and disassembly training in the automotive industry. 
The most related work is Capgemini which has developed an augmented reality-based training environment, that enables users to virtually train in the entire sequence of assembling a dashboard using a HoloLens (in use at the BMW Group) or a tablet \footnote{\url{https://www.capgemini.com/news/client-stories/augmented-reality-trains-employees-for-the-automotive-assembly-line/}}.
A VR system to teach students how to assemble and disassemble automatic vehicle transmissions \cite{farkhatdinov2009development}. 
Manuel et al. \cite{garcia2009simulation} investigated the assembly of the front components of a vehicle using VR tools. 
Becker et al. \cite{becker2011using} presented VR tools in designing the assembly and disassembly process of a new vehicle model. 
Macaria et al. \cite{hernandez2021development} developed a VR system to train students in automotive systems engineering academic programs, as a practical training-learning tool.
A VR training environment was developed to train assembly line workers in the automotive industry \cite{schwarz2020learning}. 
A VR method \cite{caputo2018use} was proposed to validate the design of workplaces on automotive assembly lines, which is applied by Fiat Chrysler Automobiles groups.
Langley et al. \cite{langley2016establishing} developed the virtual training system within the VISTRA project, using real end users from the OPEL automotive plant.
Although these studies applied virtual reality to the automotive industry, they focus on vehicle assembly rather than engine assembly/disassembly and are generally proprietary.

In addition, vehicle manufacturers such as Volkswagen, BMW, and Ford have applied VR technology to the automotive industry. 
We compare the current well-known VR systems in the automotive industry with our VR system, 
as shown in \autoref{tab:related-work}, in which only our VR system provides students with training in automotive engine assembly and disassembly.
In summary, there is a lack of VR systems that are used to train students for automotive engine assembly and disassembly.

\begin{table*}[!ht] \centering \small
\renewcommand{\arraystretch}{0.8}
\setlength\tabcolsep{4pt} 
\caption{Overview of the current VR systems in the automotive industry and our VR training system for automotive engine assembly and disassembly. }
\begin{tabular} { >{\columncolor[gray]{0.8}} l | l l c l m{7.8cm}} \toprule
Company & Objects & VR devices & Copyright & Tasks & Remarks \\ \midrule
OPEL \cite{langley2016establishing} & Vehicle & unknown & Proprietary & \makecell[l]{Vehicle assembly} & Virtual simulation and training of car assembly and service processes in digital factories. \\ \midrule
FIAT \cite{caputo2018use} & Vehicle & unknown & Proprietary & \makecell[l]{Vehicle assembly} & Workplaces on automotive assembly lines in a virtual environment. \\ \midrule
Volkswagen & Vehicle & HTC Vive & Proprietary & \makecell[l]{Vehicle assembly \\ and disassembly} & Employees have to complete tasks within a certain timeframe, such as installing a brake or wires. \\ \midrule
BMW & Vehicle & HTC Vive & Proprietary & \makecell[l]{All-round \\ VR-training} & Various applications of VR in training, such as assembly line, customer service employees, and safety managers. \\ \midrule
Peugeot & Vehicle & unknown & Proprietary & Learning by living & Training the employees for vehicle assembly in a safe, healthy and efficient environment. \\ \midrule
Audi & Vehicle & HTC Vive & Proprietary & Assembly \& logistics & Training employees by using a VR system for an efficient vehicle assembly. \\ \midrule
\textbf{Ours} & \textbf{Engine} & HTC Vive & \textbf{\makecell[l]{Free use \& \\ open source}} & \makecell[l]{Engine assembly \\ and disassembly} & Training students to learn the automotive engine assembly and disassembly with in our VR system. \\ \midrule
\end{tabular}
\label{tab:related-work}
\end{table*}

\section{Methodology}
\label{sec:methodology}
In this section, we describe the design and methodology of our VR system in detail, including system design, and experimental setup.
\subsection{System Design}
\label{subsec:System}
\begin{figure}[!ht] \centering
    \includegraphics[width=0.85\linewidth,trim={30 10 20 10},clip]{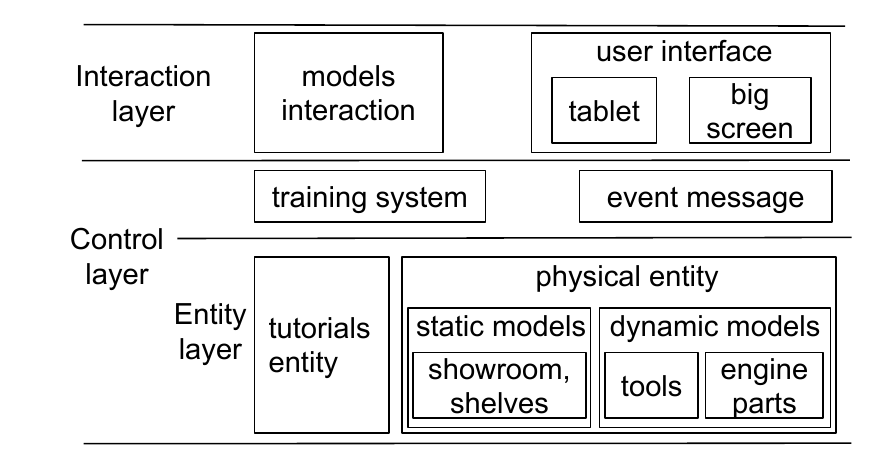}
    \caption{The multi-layer architecture of our VR system.} \label{fig:VR_system_architecture}
\end{figure}

Our VR system provides a training- and examination-mode required for student training,
in which the user interactions are designed into two independent scenarios.
Technically, these two modes contain identical operational logic for the disassembly and assembly process, that can be decoupled and implemented at the architectural levels, as shown in \autoref{fig:VR_system_architecture}.

The training mode focuses on training students in disassembly and assembly skills, which provides text tutorials and voice prompts to guide them during the practice of the specified actions. 
The examination mode focuses on evaluating students' performance.
In this mode, the VR system turns off the text tutorials and voice prompts, scores students' performance with a series of evaluation rules and finally provides a score.
our VR system records student scores locally and online, in which teachers can collect data on students' performance and improve training efficiency.


Architecturally, we design the VR training system with a bottom-up approach.
The bottom-up design provides high scalability of our VR system and enables the template-based development of immersive VR training, which significantly improves development efficiency.
Furthermore, the bottom-up design enables developers efficient teamwork by splitting the development into units and combining units flexibly.
In the entity layer, the training process is recorded into configurable data.
Next, we design the control layer by considering the demand for student training and eventually develop the interaction layer according to product requirements.
Under this architecture, the interaction in the VR system is reusable by modifying the control layer for developing products.
The entity layer can be replaced with a new training scene for developing a training task.
The multi-layer architecture significantly contributes to the development of VR systems for other tasks in the automotive industry with our open-source code. 

\subsection{Interaction Layer}
\label{subsec:interface}

As shown in \autoref{fig:VR_system_architecture}, the interaction layer is the top-level structure in our VR training system.
We split the interaction layer into two modules of model interaction and user interface. 




\begin{figure*}[!ht] \centering
    \begin{subfigure}[t]{.36\textwidth}
        \includegraphics[width=\textwidth,trim={150 100 380 100},clip]{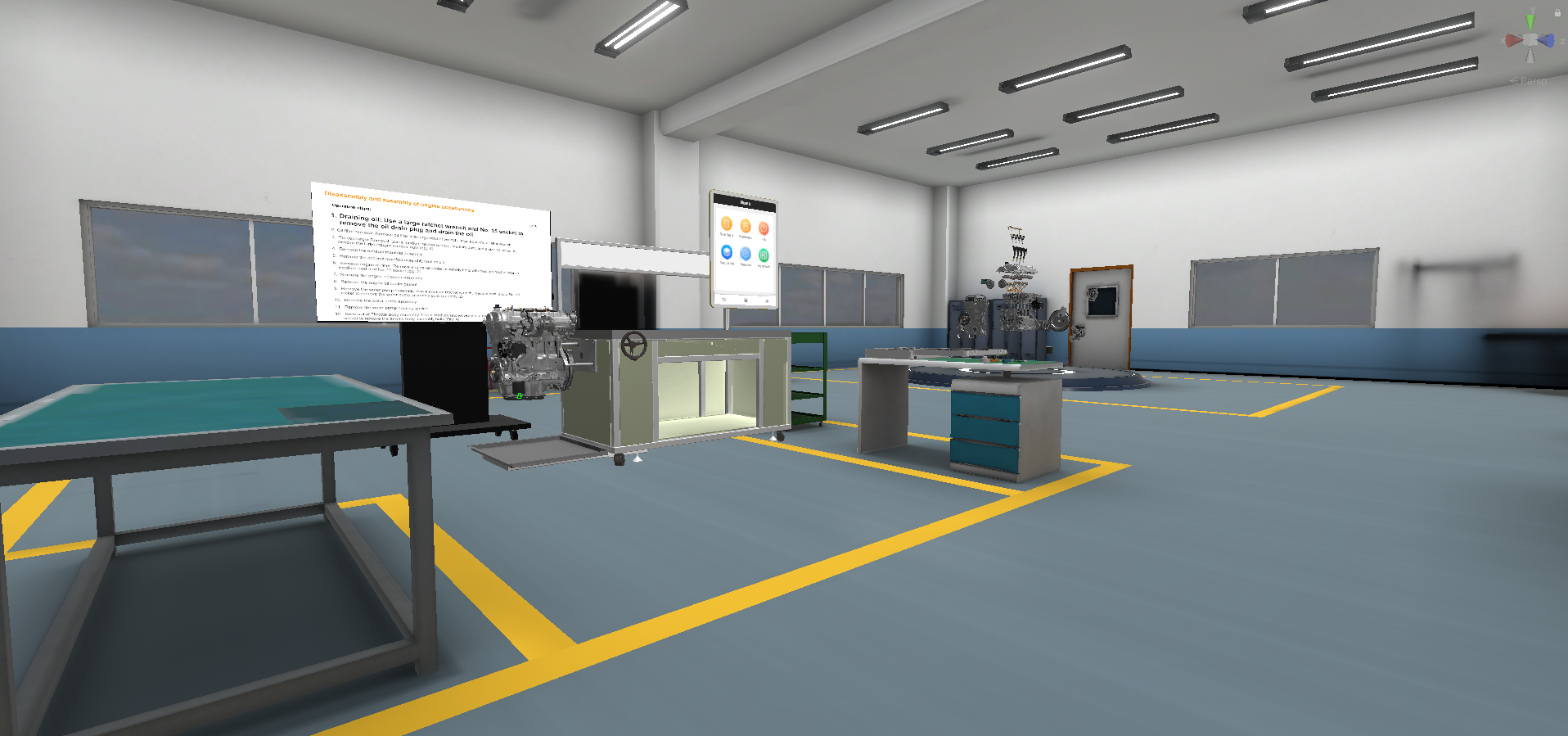}
        \caption{Virtual environment of our VR system}
        \label{fig:lab}
    \end{subfigure} \hspace{-2mm}
    \begin{subfigure}[t]{.36\textwidth}
        \includegraphics[width=\textwidth,trim={165 30 270 120},clip]{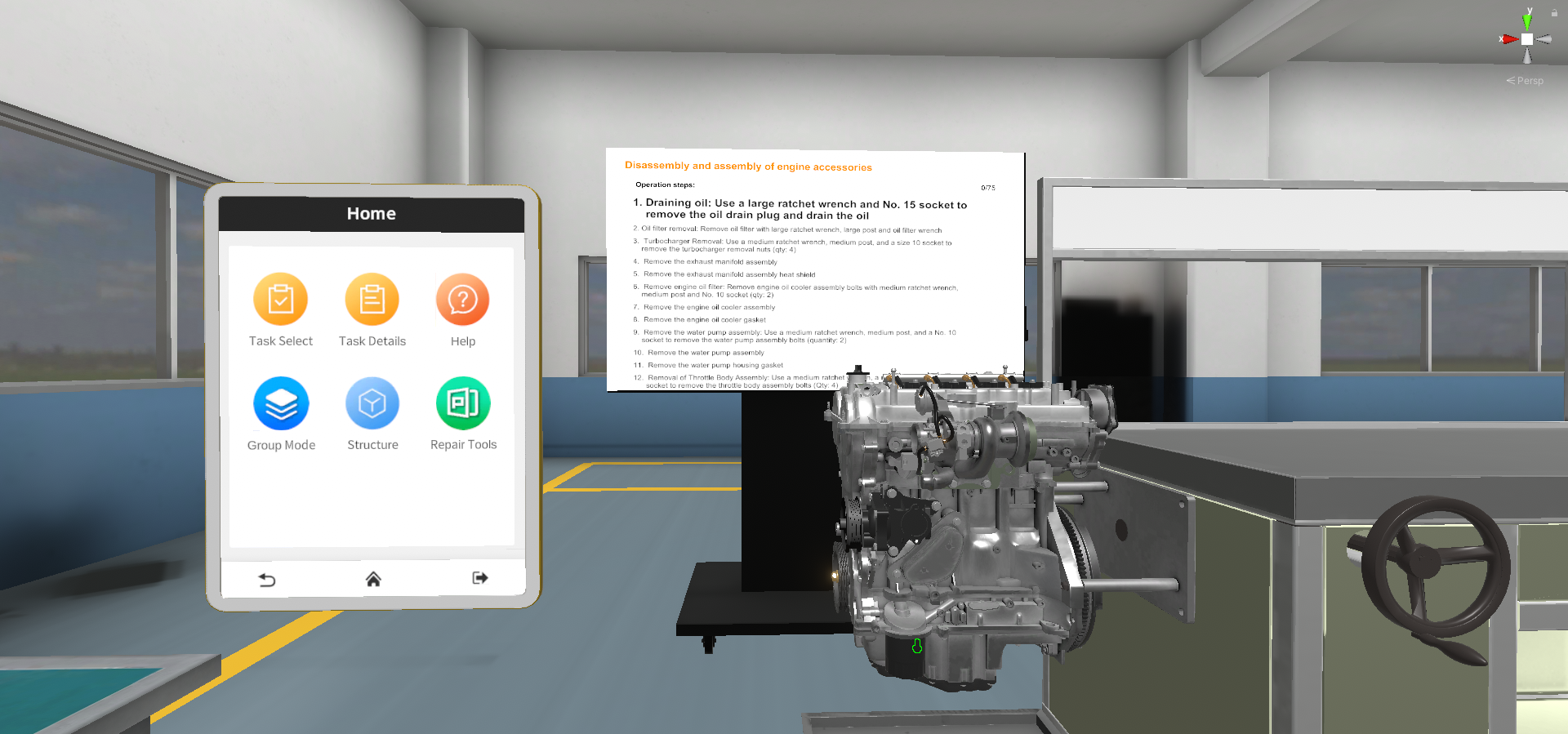}
        \caption{Virtual operation system}
        \label{fig:system}
    \end{subfigure} \hspace{-2mm}
    \begin{subfigure}[t]{.24\textwidth}
        \includegraphics[width=\textwidth,trim={70 10 0 115},clip]{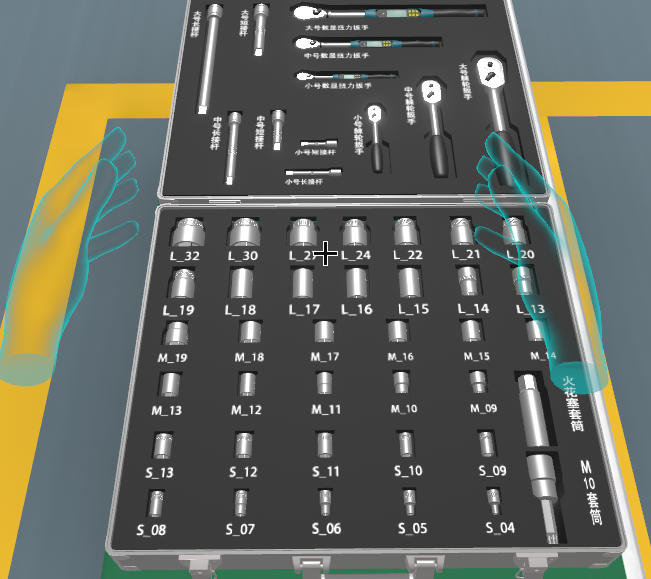}
        \caption{Virtual tools}
        \label{fig:tools}
    \end{subfigure} \\
    \begin{subfigure}[t]{.36\textwidth}
    \includegraphics[width=\textwidth,trim={200 5 350 230},clip]{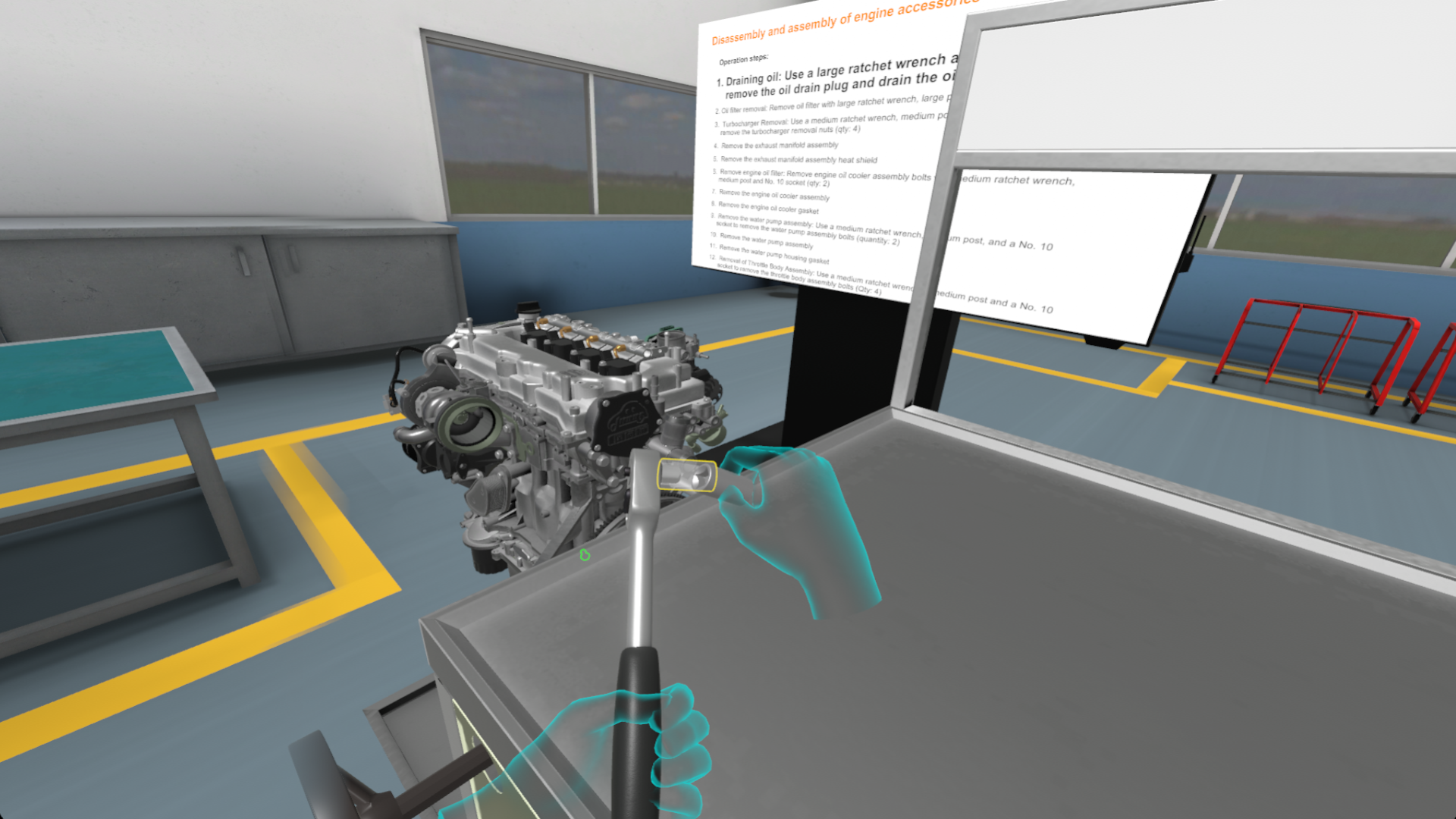}
        \caption{Disassembly operation}
        \label{fig:disassembly}
    \end{subfigure} \hspace{-2mm}
    \begin{subfigure}[t]{.36\textwidth}
        \includegraphics[width=\textwidth,trim={200 30 150 70},clip]{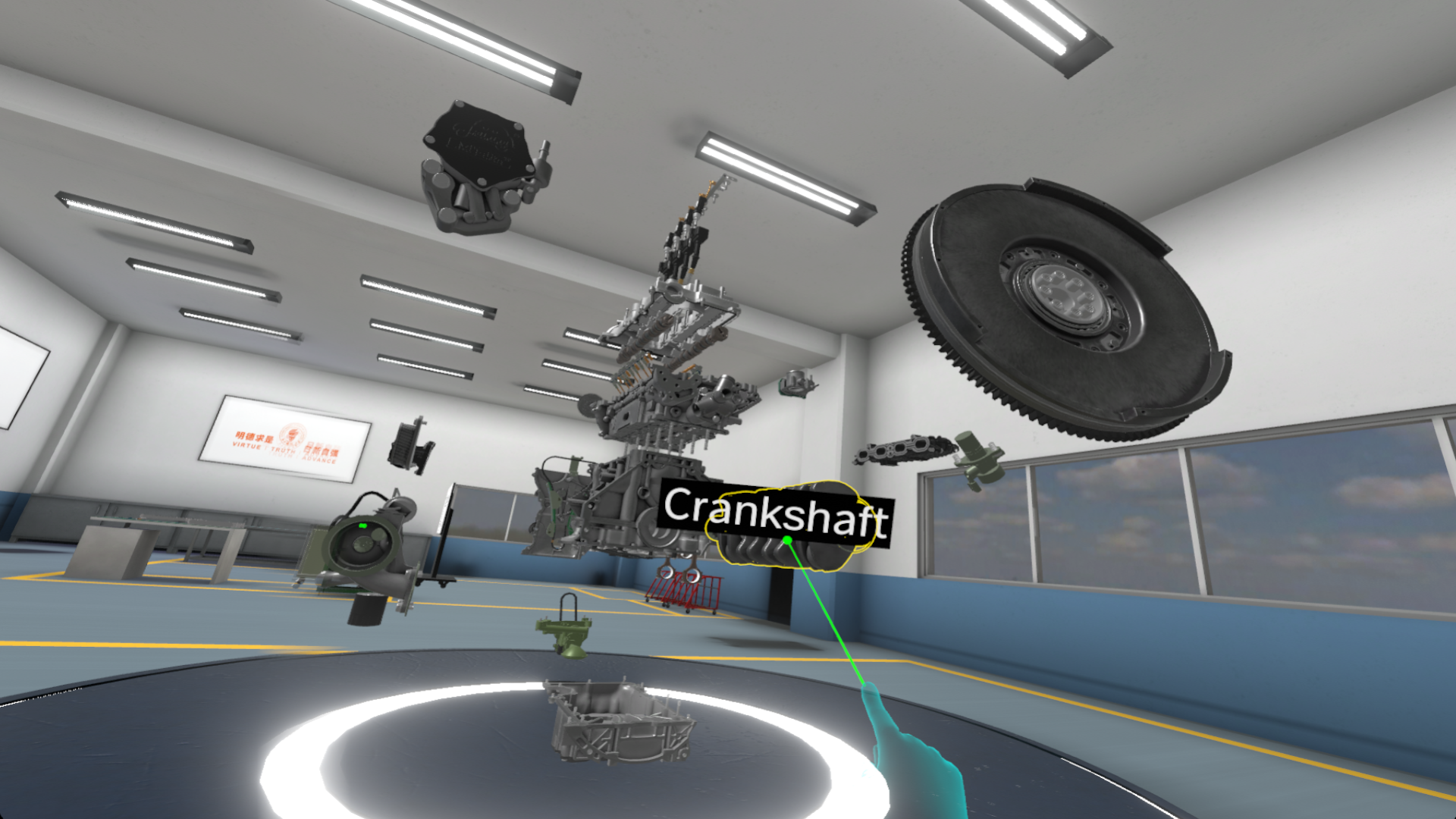}
        \caption{Exploded view and assembly}
        \label{fig:exploded_view}
    \end{subfigure} \hspace{-2mm}
    \begin{subfigure}[t]{.24\textwidth}
        \includegraphics[width=\textwidth,trim={260 140 40 10},clip]{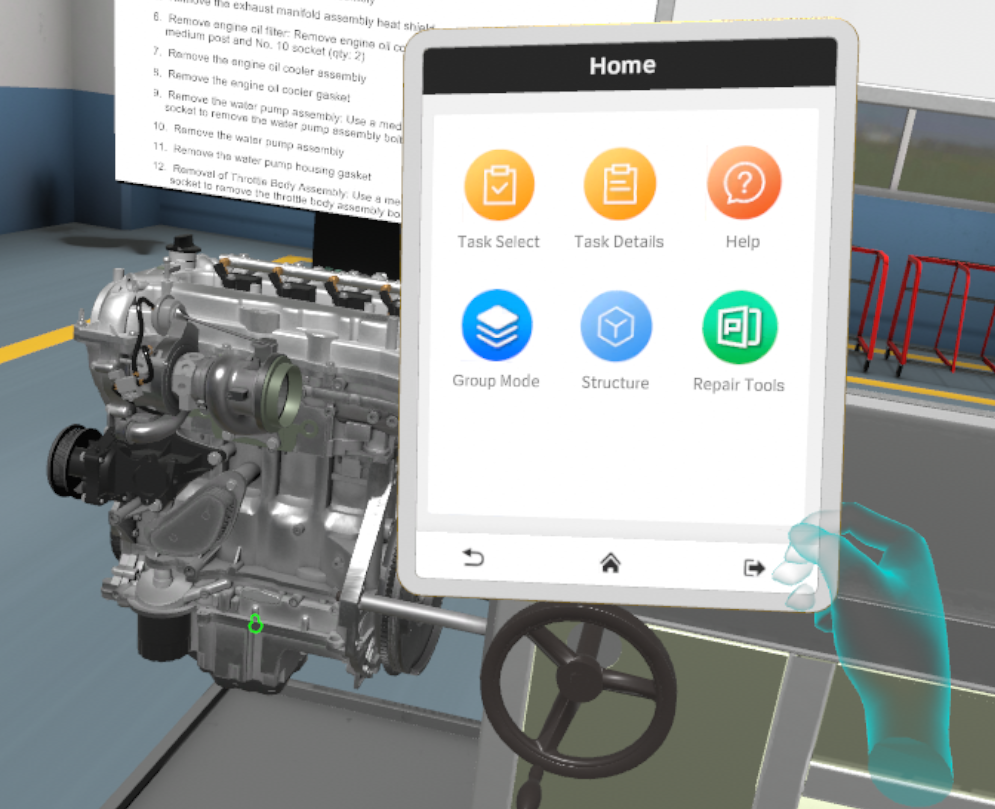}
        \caption{User interface of virtual tablet}
        \label{fig:pad_UI}
    \end{subfigure}
    \caption{Illustration of our VR system. (a) shows the virtual environment of our VR system. (b) presents the virtual operation system from the students' view during the learning. (c) displays the tools for engine assembly and disassembly in the VR system. (d) shows a scene of the disassembly operation. (e) displays the exploded view of the engine and a scene of the assembly operation. (f) displays the user interface of the Pad in our VR system. For more information about the operation of our VR system, please refer to our video description.}
    \label{fig:vr_lab}
\end{figure*}

\subsubsection{Model Interaction}
In our VR system, model interaction contains two parts of model construction and interaction. 
Model construction generally covers the 3D modelling of the VR system and its textures and materials.
We use the rendering technology provided by Unity3D for natural lustre with natural light and shadow effects in the virtual scene. 
The 3D models provide the system environment that is rendered statically in the scene, as shown in \autoref{fig:vr_lab}(a).

The models can be divided into two types non-interactive static scenes (e.g., buildings, floors, shelves) and interactive dynamic models (e.g., engine parts and tools).
In this work, we develop the model of the Buick Verano engine, as shown in \autoref{fig:vr_lab}(b) and \autoref{fig:vr_lab}(d).


An engine model is a collection of parts that are the objects of ActionScript in the VR system.
Although there are hundreds even thousands of parts in an engine, the actions in the operation of assembly and disassembly are limited.
All training actions can be decomposed into combinations of the basic actions: rotation, press, holding and hiding. 
For example, screwing is a combination of rotation and pressing that is the most frequent use of all actions.
We describe the content and order of these combinations as a configuration file by C\# script.



\subsubsection{User Interface}
\begin{figure*}  \centering
    \includegraphics[width=.98\textwidth,trim={5 5 60 5},clip]{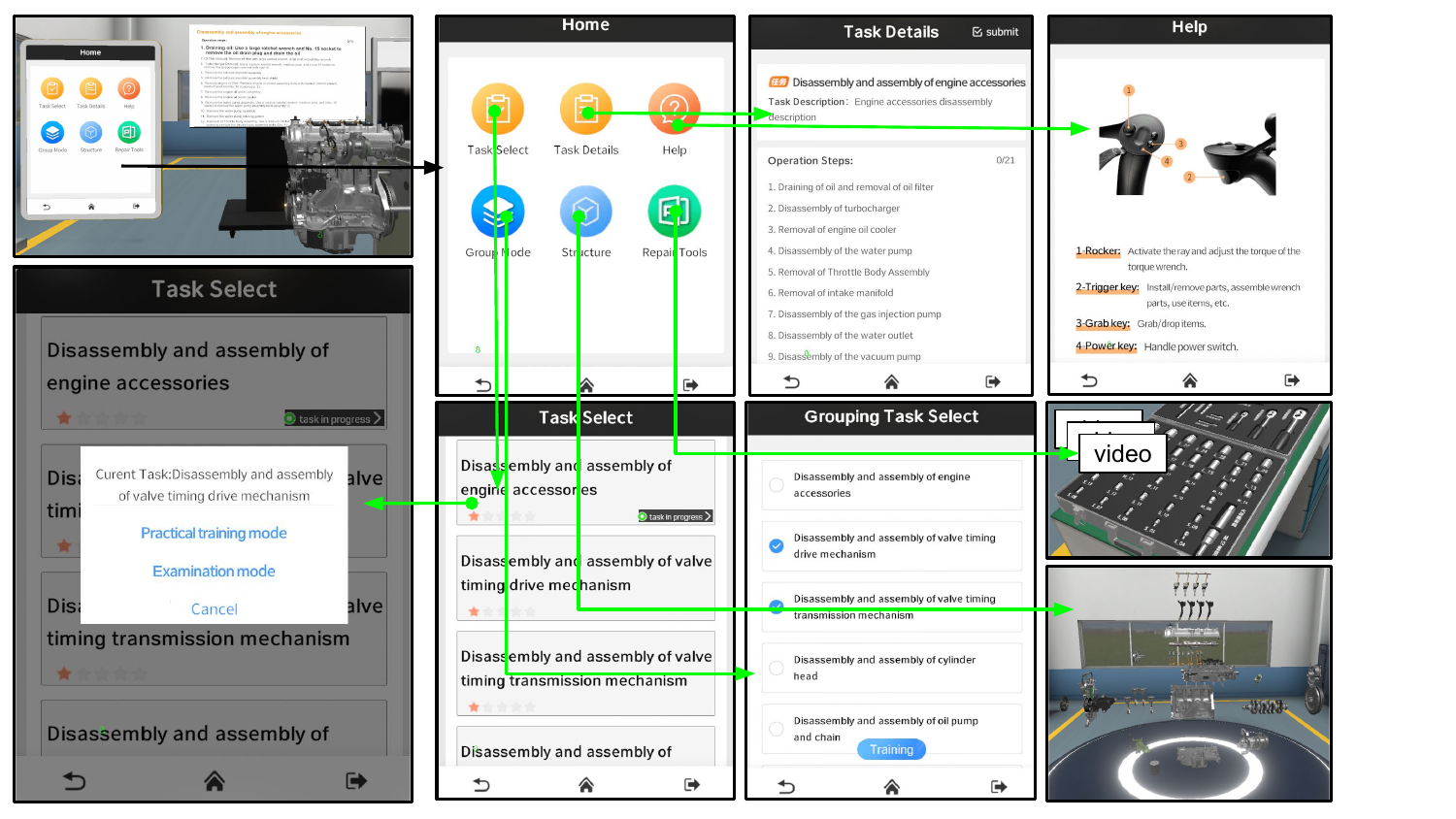}
    \caption{The user interface and operation of the virtual tablet in our VR training system.} 
    \label{fig:user_interface}
\end{figure*}

The user interface is an integral part of our VR system, 
which is integrated into a virtual tablet in the VR system, as shown in \autoref{fig:vr_lab}(f).
The user interface can be a combination of a series of interactable components (e.g., buttons, sliders, and selection boxes) and non-interactive static resources (e.g., images and description text).
In this work, we apply the Doozy UI Manager to manage and control these interactable components.
The training process of assembly and disassembly could be accessed through the tablet. 
We provide six entrances on the tablet's home page that carry the two main functional modules of training and examination.
\autoref{fig:user_interface} shows the user interface and operation of the virtual tablet in our VR system.

The training mode contains three relevant entrances: grouping mode, structure display, and tools learning. 
In the grouping mode, users can choose the chapter they wish to enter. 
The training mode provides voice prompts during the operation to guide users about how to assemble and disassemble step by step without scoring.
The other two portals present users with static resources in the form of tutorials, which do not involve interactive operations. 
In tools learning, it automatically plays an instructional video of using tools.

For the examination mode, we provide two entrances of Task Selection and Task Details. 
For Task Selection, users can enter the system and select the training of assembly or disassembly. 
The large screen shows the operations in progress when it gets access to the training of disassembly or assembly. 
In addition, students can access the Task Details module to learn all the steps of assembly and disassembly.

\subsection{Control Layer} 
Here, we address the methods of massage publishing and the training system in the control layer of our VR system.

\subsubsection{Event Message}
In general, the state of the objects in the scene needs to be updated synchronously when an operation is conducted. 
The action guide and voice prompt are refreshed when an operation is done.
In this work, we design a communication method for message publishing, which decouples message notifications and actions to improve coding efficiency.
The message system was implemented based on the Doozy UI Manager, in which the transceiver mechanism of the sender and receiver are used to process the message.
We designed the message data structure that consists of an action type and a target object.

\subsubsection{Training System}
The training system consists of three modes: training and examination. 
First, in the training mode, we provide automatic voice guidance for the learning process and 
video tutorials for students to learn the usage of various tools and basic engine disassembly and assembly skills. 
Second, we designed an examination mode to assess the student's performance in automotive engine assembly and disassembly.
In the examination mode, students complete the examination independently without any hints.
The results are recorded when a student completes a task.
The student's performance is evaluated with a score according to the scoring rules that can be predefined by teachers.


\subsection{Entity Layer} 
In our VR system, the entities contain tutorial entities and physical model entities that are replaceable and reusable for various engine models.

\subsubsection{Tutorials Entity}

Tutorial entities represent the hints that prompt students to learn automotive engine assembly and disassembly.
In this work, the tutorial guides students to practice the disassembly and assembly of the Buick Verano engine.
The tutorial entity could be replaced easily by a new tutorial when a new automotive engine is conducted in the VR system. 
Therefore, our VR system is a configurable system for the assembly and disassembly training of various automotive engines.

\subsubsection{Physical Entity} 
\begin{figure}[!ht]  \centering
    \includegraphics[width=0.9\linewidth]{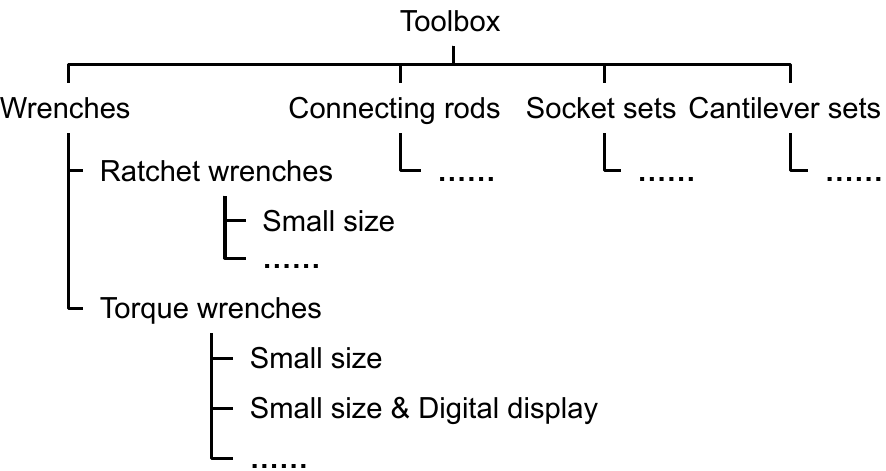}
    \caption{An illustration of the tools module in the VR system.}
    \label{fig:toolbox}
\end{figure}

The automotive engine is a model collection of physical entities.
The physical models contain static models (classroom, console, toolbox) and dynamic models (engine and tablets). 
Specifically, we technically design the tools module as the architecture shown in \autoref{fig:toolbox}.
Each of these models is a separate 3D component with the attributes of shapes, colours, and physical properties.
These models can be imported from Unity 3D and are reused for various tasks.

\section{System Testing and Optimization}
\label{sec:system_optimization}

We test the VR training system to verify its correctness and robustness, and propose solutions to optimize the VR system.
In this section, we present the testing and optimization of the VR system.

\subsection{Functional Test}
In general, we need to verify the functional integrity and correctness of the training for automotive engine disassembly and assembly.
Furthermore, it is essential to test the computing costs of the VR system. 
Specifically, we test the functional correctness of the VR system by considering the aspects below:
\begin{enumerate}
    \item the user interface on the tablet is executed to access these functions smoothly.
    \item the screen displays the completion progress correctly.
    \item the tools and models work smoothly for the assembly and disassembly.
    \item auxiliary functions such as video learning and general engine animation run smoothly.
\end{enumerate}

The user's immersive experience, video viewing experience, and interactive experience are important components of the quality of VR products.
The immersive experience is basically reflected in screen resolution and frame rate.
The screen resolution is determined by the hardware and the resolution of the 3D models.
The frame rate depends on the performance of the computing unit and the computing costs that can be optimized technically.
In general, VR products should perform at 90 fps to prevent dizziness for a smooth and non-vertiginous user experience.

\subsection{Computing Test} 
We use Unity UPR, a professional performance detection tool officially launched by Unity, to test the frame rate of our VR system.
The frame rate of the VR system would be loaded on the web page by Unity UPR.
The key performance is shown in \autoref{tab:test-data}.
\begin{table}[!ht] \centering \small
    \caption{The results of computing test by Unity UPR}\label{tab:test-data}
    \begin{tabular}{l|l}  \toprule
        Performance & Value \\ \midrule
        Average frame rate & 68.52 fps \\
        Mean frame time & 15.06 ms \\
        Peak of reserved memory & 1523.583 MB \\
        DrawCall peak & 880 ms\\ \bottomrule
    \end{tabular}
\end{table}
The average frame rate of our VR system on the computing unit is approximately 68 fps, and the memory consumption is around 1.5GB.
Although our VR system runs smoothly on the hardware system, the average frame rate of 68 fps does not reach the expectation of 90 fps.
We propose two possible technical issues:
\begin{enumerate}
    \item the Vsync module needs a lot of computing costs for the vertical synchronization switch of the program;
    \item too many DrawCalls for the rendering inside Unity takes a lot of computing costs.
\end{enumerate}
In the following, we address the optimization from these two perspectives and analyze the performance of the optimized VR system, as shown in \autoref{subsec:vsync} and \autoref{subsec:drawcall}.

\subsection{VSync optimization}
\label{subsec:vsync}

Vertical Sync, or VSync, is a graphics technology that synchronizes the refresh rate and monitor's frame rate to prevent screen tearing.
Specifically, VSync limits the output frame rate of the GPU to the same as the monitor’s refresh rate.
Although VSync provides a better gaming experience by preventing screen tearing, it limits GPU from overworking.
In this work, we optimize the VR system for a better frame rate by exploring the different options of the vertical sync technique.
There are three options for vertical sync in the Unity editor:
\begin{enumerate}
    \item Don't Sync: Do not turn on vertical synchronization, and directly let the display start drawing after the graphics card draws a picture;
    \item Every V Blank: Turn on vertical sync and synchronize on every vertical sync signal;
    \item Every Second V Blank: Turn on V-Sync and synchronize with the V-Sync signal every second.
\end{enumerate}
By default, the Unity editor selects Every V Blank, and the frame rate is completely controlled by the vertical sync signal. 
We investigate the frame rate of our VR system within these options. 
The results are shown in \autoref{tab:vs-test}.

\begin{table}[!ht] \centering \small
\setlength\tabcolsep{2pt}
        \caption{The results of the frame rate of our VR system with different V-Sync options. VS\_On means to turn on the vertical synchronization, and VS\_Off means to turn off the vertical synchronization.} \label{tab:vs-test}
        \begin{tabular}{l|ccc} \toprule
            Case & \makecell[c]{Maximum frame\\time (ms)} & \makecell[c]{Average frame\\time (ms)} & \makecell[c]{Average frame\\rate (fps)} \\ \hline
            VS\_On & 470.21 & 15.06 & 65.61 \\
            VS\_Off (ours) & 445.72 & 3.94  & \cellcolor{gray!30}\textbf{264.28} \\ \midrule
            \rowcolor{gray!30}Optimized ratio & \textbf{5.2\%} (\textbf{$\uparrow$}) & \textbf{73.8\%} (\textbf{$\uparrow$}) & \textbf{302.8\%}(\textbf{$\uparrow$}) \\ \bottomrule
        \end{tabular}
\end{table}

The results show that the frame time is reduced by turning off VSync.
Specifically, the average frame time is reduced from 15.06 seconds per frame to 3.94 seconds per frame, which is 73.8\%. 
The average frame rate is increased from 65.61 to 264.28, by 302.8\%.
Furthermore, we intuitively observed that the smoothness of VR is greatly improved in terms of user experience. 
We conclude that turning off vertical synchronization is an effective approach to optimize the frame rate of the VR system.

\subsection{DrawCall Batching optimization}
\label{subsec:drawcall}
Drawcall is another technical issue that we consider to be optimized for a better frame rate of our VR system.
Each Drawcall contains the information that the graphics API needs to draw on the screen, such as textures, shaders, and buffers.
Draw calls can be resource intensive, but often the preparation for a draw call is more resource intensive than the draw call itself.
In this work, we apply the approach of Drawcall batching to reduce the number of DrawCalls for optimizing the VR system.

Drawcall batching combines meshes that can be rendered with fewer Drawcalls in Unity. 
Batching processes multiple objects with the same material in batches. 
Unity provides two built-in draw call batching methods.
In the Unity rendering engine, batch processing is divided into dynamic and static batch processing. 
In the scene, dynamic objects can be moved and transformed.
Dynamic batch processing is automatically processed by the Unity rendering engine.
In general, Unity processes all objects as dynamic objects by default and performs dynamic batch operations.
Static batching means that objects are immovable and can be batched for the same material, but may take up more memory.
In our VR system, there are a lot of static objects that do not be moved and are deformed, which can be rendered statically with low computing cost.
We manually annotate these objects as static objects that Unity recognizes and batch them statically, thus reducing the number of draw calls. 


\begin{table}[!ht] \centering \small
    \caption{The number of DrawCalls of our VR system with the default design (dynamic objects) and the optimized design (partially static objects).} \label{fig:drawcall-test}
    \begin{tabular}{l|ccc} \toprule
        Setting & \makecell[c]{Maximum number\\of DrawCalls} &\makecell[c]{Average number\\of DrawCalls} \\ \midrule
        AllDynamic       & 880   & 796.346\\
        SetStatic (ours)      & 795 & 663.618\\ \midrule
        \rowcolor{gray!30}Optimized ratio & \textbf{9.7\%} (\textbf{$\uparrow$}) & \textbf{16.8\%} (\textbf{$\uparrow$}) \\ \bottomrule
    \end{tabular}
\end{table}

In this work, we test the performance of the VR system with different properties of the objects, i.e., static and dynamic objects. 
Specifically, we test the VR system with two designs of AllDynamic and SetStatic.
AllDynamic refers to the default design in which all objects are set to be dynamic.
SetStatic refers to the optimized VR system in which some objects are optimized to be static. 
The results are shown in \autoref{fig:drawcall-test} and show that the maximum number of DrawCalls and the average number of DrawCalls of the optimized VR system were reduced by 9.7\% and 16.8\% respectively, which is significantly improved.

\section{Experiments and Results}
\label{sec:evaluation_results}

Our VR system has been deployed to the course in Chinese colleges for the training of automotive engine assembly and disassembly by the partner of FXB Ltd..
We evaluate the VR system by observing the students' performance with the controlled experiments.
In this section, we address the details of the experiments and the results.

\subsection{Experimental Setup}
\label{subsec:exp}
\begin{figure}[!ht] \centering
    \includegraphics[width=0.48\textwidth,trim={5 50 5 20},clip]{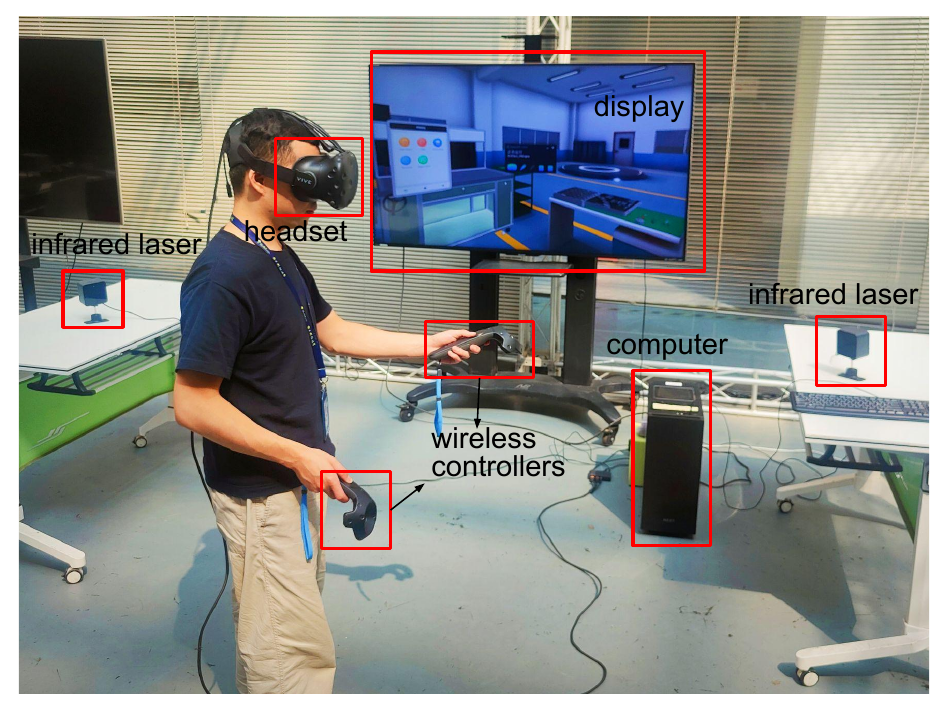}
    \caption{The hardware setup of our VR system.}
    \label{fig:setup}
\end{figure}

A VR system needs a hardware system with a computing unit and a virtual reality device. 
In this work, we use a computer equipped with an Intel 8-core i7-8700 CPU, 16G memory, and NVIDIA GeForce GTX 1080 graphics card as the computing unit.
The VR device is an HTC Vive with a virtual reality headset and accessories.
Our VR system implements a small room-scale virtual reality development, where students can freely learn automotive engine assembly and disassembly.
Multiple external base units contain an array of LED lights and two infrared lasers and are installed in the play area.
\autoref{fig:setup} shows the hardware setup of our VR system.
Note that the physical big screen shows the same view as the headset, which is essential and helpful for the evaluation by supervisors and learning of other students in the queue.

\subsection{Ablation Study}
\label{subsec:ablation}

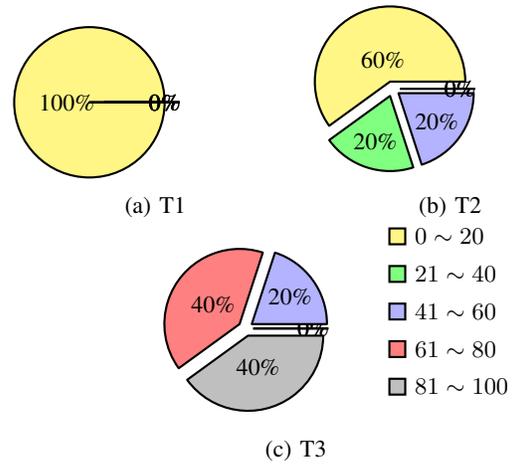
\begin{figure}[!ht] \centering \small
    \begin{subfigure}[b]{0.22\textwidth}
    \begin{tikzpicture}
    \pie[
        pos={0,5.0},
        radius=1.0,
        color = {
            yellow!60,
            green!50,
            blue!30,
            red!50,
            gray!50
        }, 
        explode = 0.1,
        text = 
    ]
    {100/$0\sim20$,
        0/$21\sim40$,
        0/$41\sim60$,
        0/$61\sim80$,
        0/$81\sim100$
    }
    \end{tikzpicture}
    \caption{T1} \label{fig:experience-a}
    \end{subfigure}
    \begin{subfigure}[b]{0.2\textwidth}
    \begin{tikzpicture}
    \pie[
    pos={3.5,3.5},
    radius=1.0,
    color = {
        yellow!60,
        green!50,
        blue!30,
        red!50,
        gray!50
    }, 
    explode = 0.1,
    text = 
    ]
    {60/$0\sim20$,
        20/$21\sim40$,
        20/$41\sim60$,
        0/$61\sim80$,
        0/$81\sim100$
    }
    \end{tikzpicture}
    \caption{T2} \label{fig:experience-b}
    \end{subfigure}
    \begin{subfigure}[b]{0.2\textwidth}
    \begin{tikzpicture}
    \pie[
    pos={0,0},
    radius=1.0,
    color = {
        yellow!60,
        green!50,
        blue!30,
        red!50,
        gray!50    
        }, 
    explode = 0.1,
    text = legend,
    ]
    {0/$0\sim20$,
        0/$21\sim40$,
        20/$41\sim60$,
        40/$61\sim80$,
        40/$81\sim100$
    }
    \end{tikzpicture}
    \caption{T3}  \label{fig:experience-c}
    \end{subfigure} \caption{The performance of the students in different experimental groups. (a) only providing voice prompts, (b) providing voice prompts, text tutorials and operation display on a virtual tablet, (c) providing voice prompts, text tutorials and operation display on a virtual tablet and screen.} \label{fig:performance}
\end{figure}

We investigate the effect of the different hints of voice prompts, text tutorials and operation display in the VR system.
For this purpose, we designed the controlled experiments with three test cases:
\begin{enumerate}
    \item T1: only providing voice prompts;
    \item T2: providing voice prompts, text tutorials and operation display on the virtual tablet;
    \item T3: providing voice prompts, text tutorials and operation display on virtual tablet and screen;
\end{enumerate}
We randomly selected 30 college students (19$\footnotesize\sim$22 years old) and randomly divided them into three groups (10 students in each group) for the three test cases and recorded their performance in the learning process.
These students have absolutely no knowledge and experience in engine disassembly and assembly.
We categorize students' performance into 5 score bands of 0$\sim$20, 21$\sim$40, 41$\sim$60, 61$\sim$80, and 81$\sim$100.
The results are shown in \autoref{fig:performance}.
The 100\% users performed with a low score of 0$\sim$20 (the yellow in \autoref{fig:performance}(a)) when the VR system only provides the voice prompts.
The 60\%, 20\%, and 20\% users scored in the range of 0$\sim$20, 21$\sim$40, and 41$\sim$60 respectively as shown in \autoref{fig:performance}(b) when the VR system provides voice prompts, text tutorials and operation display on the virtual tablet, which is much better than the user performance with only voice prompts.
Finally, the 20\%, 40\%, and 20\% users performed with the scores of 41$\sim$60, 61$\sim$80, and 81$\sim$100 respectively, as shown in \autoref{fig:performance}(c) when the VR system provides voice prompts, text tutorials and operation display on the virtual tablet and screen.
We conclude that the VR system with the hints of voice prompts, text tutorials and operation display provides users with great guidance and improvement for better learning performance.

\subsection{Usability}

\begin{figure*}[!ht] \centering \small
    \begin {tikzpicture}
        \node[draw,align = center,minimum height = 28pt] at (-5.2, 0)  (a) {S1:preparing \\ all needs};
        \node[draw,align = center,minimum height = 28pt] at (-2.5, 0)   (b) {S2:setup tools \\ (correct torque)};
        \node[draw,align = center, minimum height = 28pt] at (0.0, 0)   (c) {S3: screws};
        \node[draw,align = center, minimum height = 28pt] at (2.4, 0)   (d) {S4: oil cooler \\ \& water pump};
        \node[draw,align = center, minimum height = 28pt] at (5.1, 0)   (e) {S5: fuel rail \\ \& injectors};
        \node[draw,align = center, minimum height = 28pt] at (7.5, 0)   (f) {S6: piston \\ \& cylinder};
        \node[draw,align = center, minimum height = 28pt] at (10.0, 0)   (g) {S7:arrange \\ parts correctly};
        \draw[->,semithick] (a) node[above,xshift=1.05cm] {} -- (b);
        \draw[->,semithick] (b) node[above,xshift=1cm] {} -- (c);
        \draw[->,semithick] (c) node[above,xshift=1.55cm] {} -- (d);
        \draw[->,semithick] (d) node[above,xshift=1cm] {} -- (e);
        \draw[->,semithick] (e) node[above,xshift=1cm] {} -- (f);
        \draw[->,semithick] (f) node[above,xshift=1cm] {} -- (g);
    \end {tikzpicture}
\caption{\label{fig:operation} The seven crucial operations of engine assembly and disassembly to evaluate the effectiveness and efficiency of the VR training system.}
\end{figure*}
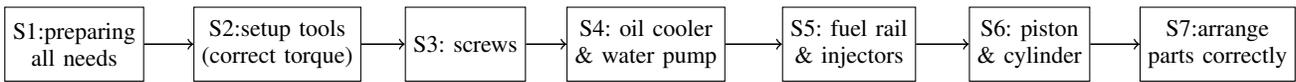

Usability evaluation focuses on how well users can learn and use a product to achieve their goals, which refers to the quality of a user's experience when interacting with products or systems in terms of effectiveness and efficiency.
In this section, we test the usability of the VR system from effectiveness and efficiency.
We evaluate the effectiveness and efficiency of the VR system by considering student performance in the seven critical operations of engine assembly and disassembly with a score, as shown in \autoref{fig:operation}.
In this section, the same total of 30 college students in the course are randomly divided into two groups (15 students in each group) to conduct the controlled experiments of traditional training and VR training.

\begin{figure}[!ht] \centering \footnotesize
    \begin{tikzpicture}
        \begin{axis}[ybar,
                enlarge x limits=.15,
                bar width=.3cm,
                legend style={at={(0.5,1)},anchor=north,legend columns=-1},
                symbolic x coords={S1, S2, S3, S4, S5, S6, S7},
                width=.5\textwidth, height=.4\textwidth,
                ylabel={Score (correctness rate)},
                ylabel near ticks,
                ymin=0, ymax=125,
                ytick={0, 20, ..., 100},
                xtick=data,
                ]
            \addplot [ybar, nodes near coords, fill=blue] coordinates{(S1,61.54)(S2,53.85)(S3,53.85)(S4,46.15)(S5,69.23)(S6,38.46)(S7,15.38)};
            \addplot [ybar, nodes near coords, fill=red] coordinates{(S1,100.0)(S2,84.62)(S3,74.92)(S4,100.0)(S5,100.0)(S6,92.31)(S7,92.31)};
            \addplot[draw=blue,ultra thick,smooth] coordinates{(S1,61.54)(S2,53.85)(S3,53.85)(S4,46.15)(S5,69.23)(S6,38.46)(S7,15.38)};
            \addplot[draw=red,ultra thick,smooth] coordinates{(S1,100.0)(S2,84.62)(S3,74.92)(S4,100.0)(S5,100.0)(S6,92.31)(S7,92.31)};
            \legend{Traditional group, VR group}
        \end{axis}
    \end{tikzpicture}
    \caption{The effectiveness of the VR system, correctness rate of the traditional group and VR group in the seven operations.}
    \label{fig:comparison}
\end{figure}
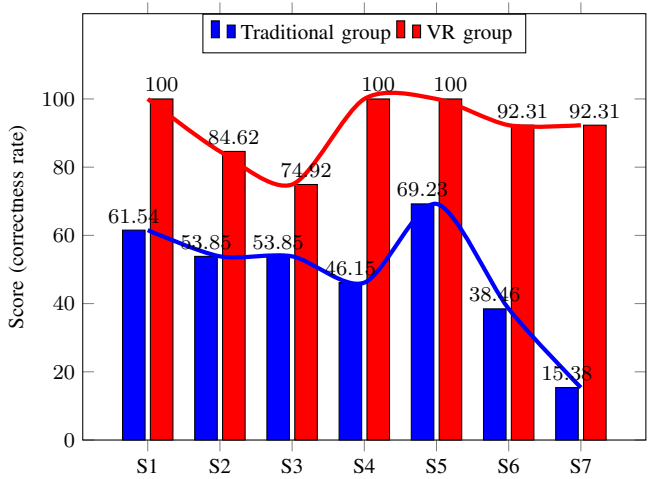

First, we observed the performance of the two groups of students in the seven important operations of engine assembly and disassembly to evaluate the effectiveness of the VR system.
Traditional training refers to the lecture in which teachers give presentations with slices.
The first group of students learn automotive engine assembly and disassembly in the traditional lecture, called the traditional group. 
The second group of students are trained with our VR system, called the VR group.
We count the correctness and calculate the score (correctness rate) for two groups of students on all seven operations, which indicates the effectiveness of the VR system.
The correctness rate is the number of correctness divided by the total number of students in each group.
The correctness rate of the traditional group and VR group are shown in \autoref{fig:comparison}.
Most of the students in the VR group finished all seven operations correctly.
Specifically, 100\% students in the VR group finished the operations of S1, S4, and S5 correctly.
Even in the worst case of operations S3, more than 70\% of the students in the VR group assembled and disassembled the engine part correctly.
However, in the best case of the traditional group, 69.23\% of students completed operation S5 correctly, and only 15.38\% of students completed operation S7 correctly in the worst case.
We conclude that the students in the VR group performed much better than the students in the traditional group for all seven operations in automotive engine assembly and disassembly.
Furthermore, teachers in the traditional group take a lot of time to avoid safety accidents that generally do not have to be considered in the VR group.
In addition, we observed that the students in the VR group showed higher motivation and interest than the students in the traditional group for learning automotive engine assembly and disassembly.
The students in the VR group took less time 
than the traditional group.  
The training in the VR group is more effective than that of the traditional group.

\begin{figure}[!ht] \centering
    \includegraphics[width=0.99\linewidth,trim={5 5 8 5},clip]{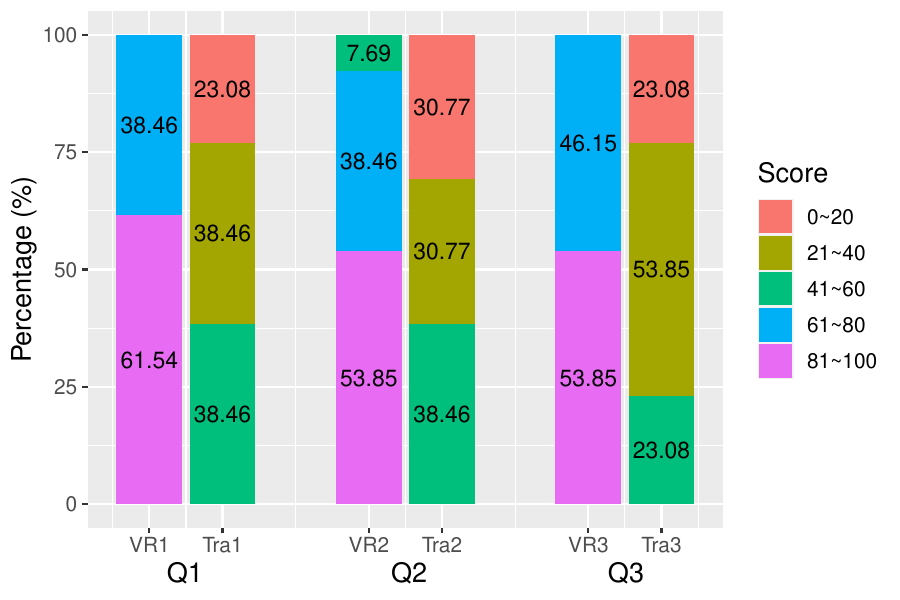}
    \caption{The comparison of efficiency for the learning experience of VR group (VR(num)) and traditional group (Tra(num)). The higher percentage of high scores, the better.}
    \label{fig:comparison2}
\end{figure}

Second, we evaluate the efficiency of the VR system by considering the proficiency and motivation of two groups of students with a score from the three aspects as below.
\begin{enumerate}
    \item Q1: the proficiency that students disassemble and assemble the automotive engine.
    \item Q2: the motivation that students learn the engine disassembly and assembly by the VR system.
    \item Q3: the proficiency that students use tools.
\end{enumerate}
We divide their scores into five intervals of 0$\sim$20, 21$\sim$40, 41$\sim$60, 61$\sim$80, and 81$\sim$100.
The percentage of the scores in the five intervals from the three aspects are shown in \autoref{fig:comparison2}. 
The higher percentage of high scores indicates that the VR system performs higher efficiency.
The results show that the students in the VR group significantly outperform the students in the traditional group.
Most students in the VR group obtained a score of more than $60$.
Conversely, all students in the traditional group scored less than $60$.
For example, 61.54\% and 38.46\% students in the VR group scored 81$\sim$100 and 61$\sim$80 respectively for the evaluation Q1.
However, none of the students in the traditional group scored more than $60$ for the evaluation Q1.
For the evaluation Q3, 53.85\% and 46.15\% students in the VR group performed with a score of 81$\sim$100 and 61$\sim$80 respectively.
The students in the VR group showed similar results with their performance in the evaluations Q1 and Q3.
We conclude that the VR system is efficient for training in automotive engine assembly and disassembly.

\section{Discussion and Conclusions}
\label{sec:discussion_conclusion}
In this section, we discuss the interesting issues, conclude this work and outlook future work.

\subsection{Discussion}
\label{subsec:discussion}


First, 
the VR system with VSync optimization provides a remarkable immersive experience with 264.28 FPS on the computer with an NVIDIA GTX 1080 GPU.
However, as no free lunch theorem, VSync optimization is technically controversial since it may cause "screen tearing" generally.
From the experimental results, our VR training system has not suffered from this problem on the computers with the buffer mechanism, 
as shown in the video description.
In addition, we empirically speculate that VSync optimization is suitable for the case of moderate screen switching in our VR training system.
Furthermore, the high frame rate of our VR system is crucial to fit the various users' computers of different computing power, which is a challenge for a VR system.

Second, the results indicate that different interaction schemes achieve different improvements for training students.
In the VR system, we integrated three interactions of voice, text, and operation display on the virtual tablet and screen. 
As the results in \autoref{fig:performance}, 
100\% students obtained a low score of 0$\sim$20 when they were only provided voice prompts.
In contrast, 80\% students performed with the scores of 61$\sim$100 and no students showed the performance with a low score of 0$\sim$20 in the group of screen interaction, which significantly outperforms the students in the other two groups.
We suspect that students hardly understand the expected operation during the training process with only voice prompts and tend to learn skills by watching in the virtual system like learning in the real world.
Therefore, a crucial method to improve the VR training system is to realistically simulate the operation in the real world.

Third, the elephant in the room for the VR training systems is the reality gap between virtual learning and learning in the real world.
We noticed that students' performance varied for different scores in different operations as shown in \autoref{fig:comparison}.
Students in the VR group had an average correctness rate of $74.92\%$ in operation S3 which is significantly worse than the performance in the other operations.
We speculate that the reality gap is one of the main reasons for this issue.
The VR training for operation S3 may need to be improved in future work.
Realistic 3D models and reasonable learning processes are effective methods to mitigate the reality gap of the VR training system.
Furthermore, we noticed that operation S7 is difficult to learn for students and only $15.38\%$ students in the traditional group performed correctly. 
In contrast, the students in the VR group accomplished well with a score of $92.31$.

Finally, although the results in \autoref{fig:comparison} show the objective effectiveness evaluation (correctness rate) for the VR system, the efficiency evaluation by supervisors in \autoref{fig:comparison2} may not be objective.
We tried to design and apply a user experience survey for students to evaluate the usability, but we noticed that different students adopted significant subjective and fluctuating evaluation criteria when they scored their experience for reporting the survey. 
For example, a few students performed poorly and negatively in the training with the VR system, but reported a high score for their experience in the survey.
The current efficiency evaluation by the supervisors is a good balance between reasonability and objectivity.
Therefore, we conclude that despite the evaluation by the supervisors may not be objective, the efficiency evaluation by the supervisors is a reasonable and objective method to evaluate the usability of the VR training system.
Last, 
although more users in the controlled experiments will provide more accurate results, we noticed that the experimental results with 30 students in this work are sufficient to draw a convergent conclusion.

\subsection{Conclusions}
\label{subsec:conclusions}

In this work, we developed a VR system to train students to learn automotive engine (Buick Verano) assembly and disassembly by using Unity3D and VRTK plugins.
We designed the VR system with a distinct multi-layer architecture which was developed modularly.
The VR system contains replaceable engine parts and reusable tools, friendly user interfaces and guidance, and bottom-up designed multi-layer architecture, which can be extended to various engine models.
Our VR system integrates friendly and helpful guidance of voice prompts, text tutorials and operation displays to provide students with efficient, safe, low-cost training in automotive engine assembly and disassembly. 

Furthermore, we monitor and investigate the computing cost of the VR system by Unity UPR, and improve the user experience by optimizing the frame rate of the VR system.
Specifically, we reduce computing costs by optimizing the vertical synchronization to increase the frame rate and apply the DrawCall batching technology to reduce the number of DrawCall for the higher frame rate.
Our VR system ultimately reaches the requirement of 90 fps on the hardware (\autoref{subsec:exp}) for a smooth and non-vertiginous user experience.

Finally, the VR system has been implemented and demonstrated at Chinese colleges by the partner (FXB Ltd.) to train students in automotive engine assembly and disassembly.
We divided the students into traditional training and VR training to compare their learning performance.
The results demonstrate that our VR training system provides great usability (effectiveness and efficiency).
In terms of effectiveness, students in the VR training obtained significantly higher scores than the students in the traditional training.
In terms of efficiency, students in the VR training showed better proficiency and motivation than students in the traditional training.
We conclude that this work provides students with an effective and efficient VR training system for automotive engine assembly and disassembly.

We provide a free-to-use executable file (Microsoft Windows) that can be freely and easily used to implement a VR training system by those who are interested. 
Furthermore, we publicly release the source code of the VR system that can be extended to develop various VR systems in the automotive industry.
Our VR system with the released executable file and open source code significantly contributes to facilitate the development of VR in the automotive industry, particularly automotive engine assembly and disassembly.

In the future, we will improve the VR system by considering the feedback from the students and supervisors. 
Furthermore, we will apply virtual reality technology to other educational tasks in the automotive industry, for example, the education of perception systems in autonomous driving. 


\section*{Acknowledgements}


The authors would like to thank the students (Jiayu Zhang, Zhuochen Xiong, Shengdi Gao, Zhiyuan Han, etc) and the partner of FXB Ltd. (Mr Junwei Yang, etc) for their business management, valuable comments, helpful suggestions, and great support for experimentation.
In addition, the authors would like to thank 
Dr. Changchun Zhong from the University of Chicago, Dr. Fuda van Diggelen from VU University Amsterdam, and Prof. Dr. Geoff Nitschke from the University of Cape Town for their valuable comments and helpful suggestions in English writing.






\bibliographystyle{IEEEtran}
\bibliography{bibliography}

\vspace{-10mm}
\begin{IEEEbiography}
    [{\includegraphics[width=1in,height=1.25in,trim={0 0 0 0}, clip, keepaspectratio]{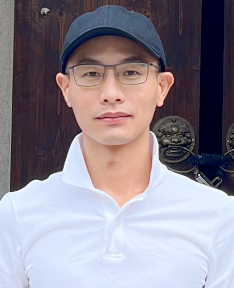}}]
    {Gongjin Lan} received his PhD in Artificial Intelligence at VU University Amsterdam in 2020. 
    He is currently a Research Assistant Professor at the Department of Computer Science and Engineering, Southern University of Science and Technology, Shenzhen, China. He has published over 20 academic papers in top international journals and conferences, including IEEE THMS, Swarm and Evolutionary Computation, Scientific Reports, PPSN. He serves as a reviewer for IEEE TEVC, Swarm and Evolutionary Computation, Scientific Reports, PLOS ONE, ICRA, IROS, GECCO, CEC, etc. His current research interests include Autonomous Driving, Evolutionary Intelligence, and AI in Health. 
\end{IEEEbiography}

\vspace{-13mm}
\begin{IEEEbiography}
    [{\includegraphics[width=1in,height=1.25in,trim={0 10 0 10},clip,keepaspectratio]{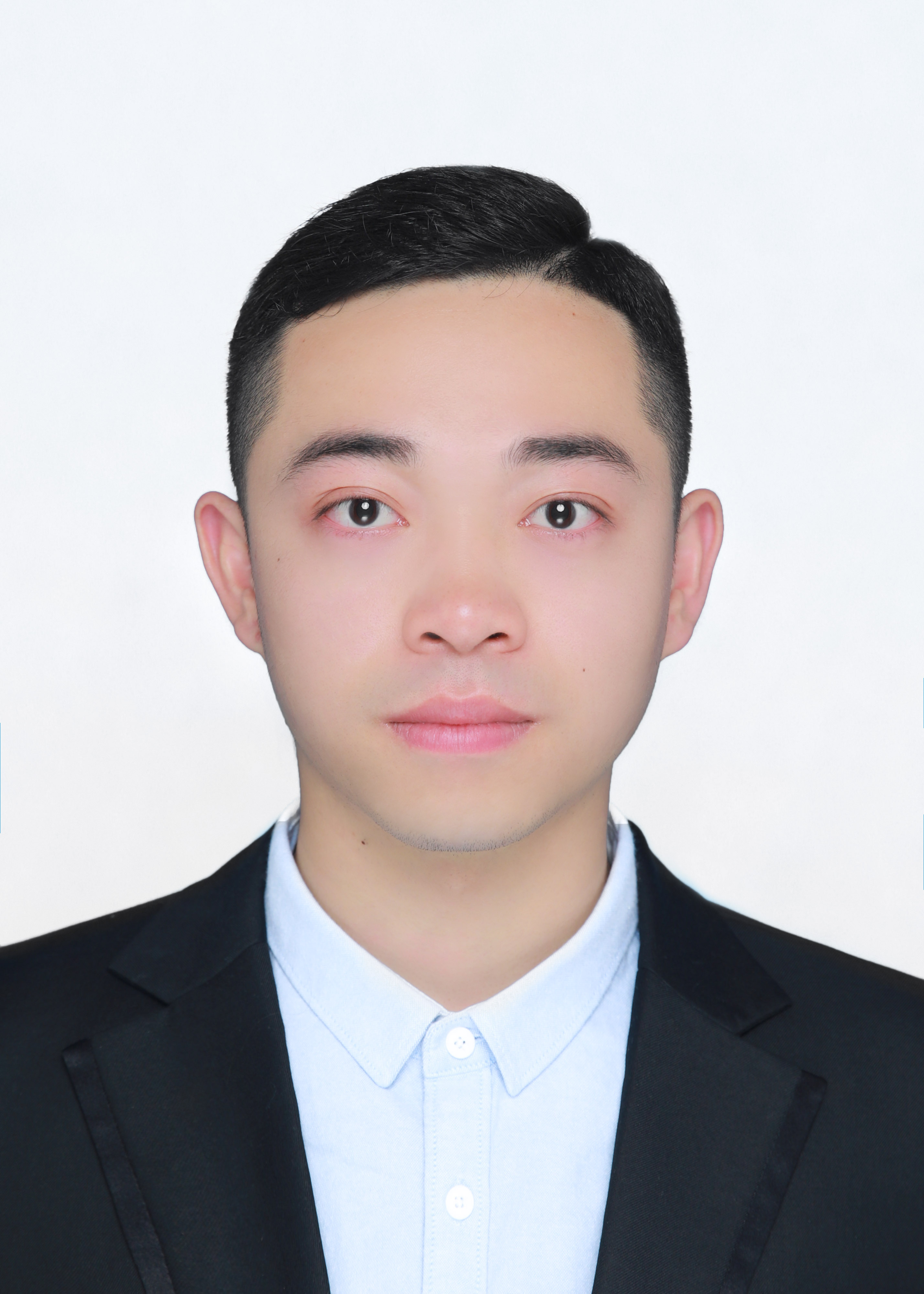}}] 
    {Qiangqiang Lai} received the B.E. degree in Vehicle Engineering from the Southwest University, Chongqing, China, in 2019. He is currently a research assistant at the Department of Computer Science and Engineering, the Southern University of Science and Technology, Shenzhen, China. His research interests include Unmanned autonomous systems, Autonomous driving, Intelligent Robots, and 3D modelling.
\end{IEEEbiography}

\vspace{-13mm}
\begin{IEEEbiography}
    [{\includegraphics[width=1in,height=1.25in,trim={0 10 0 5},clip,keepaspectratio]{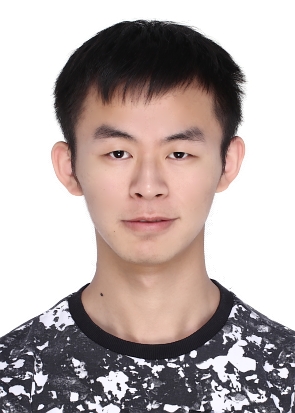}}] 
    {Bai Bing} received a B.E. degree from the Department of Computer Science and Technology, the Southern University of Science and Technology (SUSTech) in 2020. 
    He is currently a research assistant at the Research Institute of Trustworthy Autonomous Systems and the Department of Computer Science and Engineering, the Southern University of Science and Technology, Shenzhen, China. 
    His research interests include Autonomous driving, Traffic flow simulation, 3D modelling, and Robotics.
\end{IEEEbiography}

\vspace{-13mm}
\begin{IEEEbiography}
    [{\includegraphics[width=1in,height=1.25in,trim={7 10 7 10},clip,keepaspectratio]{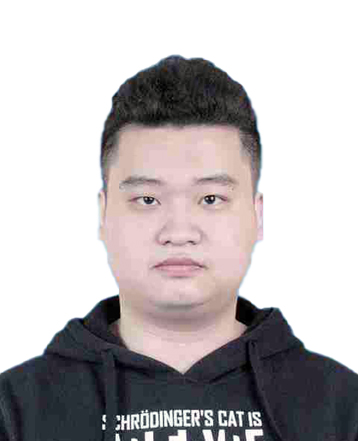}}] 
    {Zirui Zhao} received a B.E. degree in Computer Science and Technology from the Southern University of Science and Technology (SUSTech), Shenzhen, China in 2020. He is currently a research assistant at the Research Institute of Trustworthy Autonomous Systems and the Department of Computer Science and Engineering, SUSTech, Shenzhen, China. His current research interests include Autonomous driving simulators, Robotics, and Software engineering.
\end{IEEEbiography}

\vspace{-13mm}
\begin{IEEEbiography}
    [{\includegraphics[width=1in,height=1.25in,trim={0 5 0 15},clip,keepaspectratio]{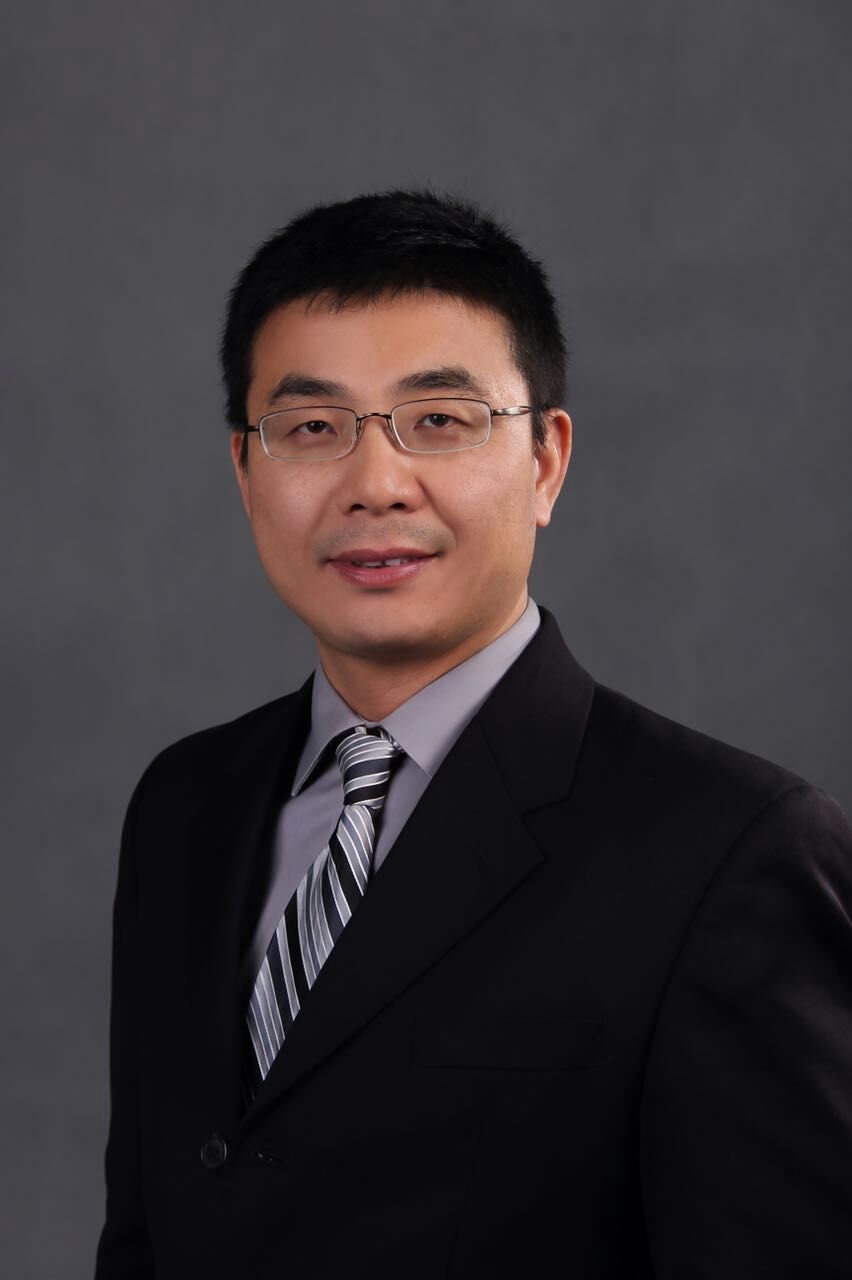}}]
    {Qi Hao} received the B.E. and M.E. degrees from Shanghai Jiao Tong University, Shanghai, China, in 1994 and 1997, respectively, and the Ph.D. degree from Duke University, Durham, NC, USA, in 2006, all in electrical and computer engineering. He is currently an Associate Professor at the Southern University of Science and Technology, Shenzhen, China. He was a Postdoctoral Fellow with the Center for Visualization and Virtual Environment, University of Kentucky, Lexington, KY, USA, focusing on 3D computer vision. From 2007 to 2014, he was an Assistant Professor at the Department of Electrical and Computer Engineering, University of Alabama, Tuscaloosa, AL, USA. His current research interests include Intelligent unmanned systems, Machine learning, and Smart sensors.
\end{IEEEbiography}

\end{document}